\documentclass[]{raa}            % referee version: for submission

\usepackage{natbib}
\usepackage{booktabs,threeparttable}
\usepackage{multirow}
\usepackage{longtable}
\usepackage{graphicx,times}             %for PS/EPS graphics inclusion, new

\providecommand{\bm}[1]{\mbox{\boldmath$#1$\unboldmath}}

\begin{document}

   \title{Dependence of low redshift Type Ia Supernovae luminosities on host galaxies}
% $^*$
%\footnotetext{\small $*$ Supported by the National Natural Science Foundation of China.}

 \volnopage{ {\bf 201X} Vol.\ {\bf X} No. {\bf XX}, 000--000}

   \setcounter{page}{1}
   \author{Wen-Ke Liang
      \inst{1,2,3}
   \and Jian-Cheng Wang
      \inst{1,2}
   }

 \institute{Yunnan Observatory, Chinese Academy
of Sciences,  Kunming 650011, China; {\it lwk@ynao.ac.cn}\\
%        \and
%             Full institute address for the second author
        \and
             Key Laboratory for the Structure and Evolution of Celestial Objects,
Chinese Academy of Sciences,  Kunming 650011, China
        \and
        University of Chinese Academy of Sciences, Beijing, China
%\vs \no
%   {\small Received [year] [month] [day]; accepted [year] [month] [day] }
}

%% non-output

%% Abstract
\abstract{We study the relation of Type Ia Supernovae (SNe Ia) with host galaxies from a large low redshift sample. By examining the Hubble residuals of the entire sample from  the best-fit cosmology, we show that SNe Ia in passive hosts are brighter than those in star-forming hosts after light curve correction at 2.1$\sigma$ confidence level.
We find that SNe Ia in high luminosity hosts are brighter after light-curve correction at $>$3$\sigma$ confidence level. We also find that SNe Ia in large size galaxies are brighter after light-curve correction at $\geq$2$\sigma$ confidence level.
We demonstrate that the residual linearly depends on host luminosity at a confidence of 4$\sigma$ or host size at a confidence of 3.3$\sigma$.
  }

   \authorrunning{Liang, Wang}            %author_head in even pages
   \titlerunning{Dependence of low redshift Type Ia Supernova properties on host galaxies}  % title_head in odd pages
   \maketitle

\section{INTRODUCTION}           %% first-level sections will be auto-capitalized
\label{sect:1}

The use of Type Ia supernovae (SNe Ia) as standard candles in estimating cosmological distances has been proven indispensable for modern cosmology, leading to the remarkable discovery that the expansion of the Universe is accelerating \citep{1998AJ....116.1009R,1999ApJ...517..565P,2009ApJS..185...32K,2010A&A...523A...7G, 2012ApJ...746...85S}. According to the current theory, the progenitor of a SN Ia is a carbon-oxygen white dwarf that approaches the Chandrasekhar limit, resulting in a thermonuclear explosion \citep{2000ARA&A..38..191H,2012NewAR..56..122W}. However, the exact mechanism how the progenitor accumulates this mass remains uncertain. Investigations of the physical properties of SN Ia host galaxies can provide the insight for the environment of SN Ia progenitor system. Furthermore, although SNe Ia are remarkably standardizable, the corrections for light-curve width and color still result in a scatter in peak brightness of $\sim$ 0.15 mag \citep{2007A&A...466...11G,2007ApJ...659..122J,2008ApJ...681..482C}. The search for the relation of SN Ia luminosity with host galaxies will help us to reveal the origin of this scatter.

Over the years, several correlations between SNe Ia and the properties of their host galaxies have been discovered. The characteristics, such as the morphology, color, star formation rate, metallicity, and stellar age of host galaxies provide us clues to understand the progenitors. SNe Ia in E/S0 galaxies are brighter than those in later-type galaxies after light-curve shape and color corrections \citep{2009ApJ...700.1097H}. SNe Ia are brighter in massive hosts and hosts with low star formation rate per stellar mass (specific SFR) after SN Ia maximum brightness being corrected by using their light-curve shape and color \citep{2010MNRAS.406..782S,2010ApJ...715..743K,2010ApJ...722..566L}. \citet{2011ApJ...740...92G} also found that over-luminous SNe Ia tend to occur in older stellar populations after light-curve correction.

However, there still a few research about the correlations between SNe Ia and the properties of their host galaxies at low redshift. In the paper, we investigate correlations between SNe Ia and the properties of their host galaxies at low redshift. If the correlations are identified at a low redshift, host properties such as type, size, and luminosity could be combined with light curve parameters to make a further improvement for luminosity distance estimates.

In the paper, we try to study the relation of SNe Ia luminosities with host galaxy properties, such as type, size and luminosity, by using a larger low redshift sample. In Section 2, we introduce the SN Ia and host galaxy sample and outline the details of our analysis using public light curve fitting procedures SALT2 \citep{2007A&A...466...11G}. We investigate how the light-curve widths and colors of
SNe Ia vary with host galaxy properties in Section 3. In Section 4, we present the relation of Hubble residuals with light curve parameters, host
galaxy type, luminosity and size. The conclusions are presented in Section 5. Throughout we use a flat $\Lambda$CDM cosmological model with $\Omega_{M}$ = 0.270 and H$_{0}$ = 70 km s$^{-1}$ Mpc$^{-1}$.

\section{Sample and data}\label{sect:2}
\subsection{Low redshift SN Ia sample}\label{sect:2.1}

To seek the relationship between SNe Ia and host galaxies, we just choose the low redshift sample. It includes 8 main samples: Cal\'an/Tololo \citep[][29 SNe Ia]{1996AJ....112.2408H}, CfAI \citep[][22 SNe Ia]{1999AJ....117..707R}, CfAII \citep[][44 SNe Ia]{2006AJ....131..527J}, CfAIII \citep[][185 SNe Ia]{2009ApJ...700..331H}, LOSS \citep[][165 SNe Ia]{2010ApJS..190..418G} and CSP \citep[][35 SNe Ia]{2010AJ....139..519C}, CSPII \citep[][50 SNe Ia]{2011AJ....142..156S}, CfAIV \citep[][94 SNe Ia]{2012ApJS..200...12H}.
For the Cal$\acute{a}$n/Tololo, CfAI, CfAII and LOSS samples, the data were transformed by the authors from the natural instrumental system into the \citet{1992AJ....104..340L} system using linear transformations derived from stars in a limited color range. For the CfAIII, CSP, CSPII and CfAIV samples, natural system photometry is used in our analysis, and we abandon the U band because of its relatively large error.

\subsection{SN Ia Selection}\label{sect:2.2}

We exclude 150 repeated SNe Ia and 29 known peculiar SNe Ia by hand, such as SN 2000cx.
Because of the potential issue of a discontinuous step in the local expansion rate (Hubble bubble) detected by \citet{2007ApJ...659..122J}, we follow the choice of the cut at $z= 0.010$ \citep{2011ApJS..192....1C}.

For the reliable cut of light-curve data, we will select SNe Ia according to the following requirements
from the available phases $\tau = (T_{obs}-T_{max})/(1+z)$ of photometric observations, where $T_{obs}$ is the date of observing light-curve data, and $T_{max}$ is the date of light maximum:\\
(i) Measurements at five different epochs or more in the range of $\tau < +60$ days.\\
(ii) At least two measurement in the range of $\tau < +6$ days.\\

We abandon SNe Ia without reliable light-curve parameters by SALT2 light-curve fitter.
The Galactic reddening along the line of sight should satisfy $E\left(B - V\right)_{MW} < 0.5$ mag because the assumed Galactic value of $R_V = 3.1$ could not be appropriate for highly extinguished objects.  Next, we take the stretch parameter to be $-4 < x_{1} < 3$ for SALT2 and the color to be $-0.2 < c < 0.4$.
We exclude SN 2008cm and SN 2006bd due to $3\sigma$ intrinsic luminosity dispersions in the best-fit cosmology. The data of SNe Ia can be obtained on line. The results of all these selections are shown in Table \ref{sampletab}.

\begin{table*}
\caption[]{The selection of SNe Ia from low redshift samples.}
  \label{Tab:poptical}
\begin{center}
\begin{tabular}{lccccccc}
  \hline\noalign{\smallskip}
Sample& Initial$^{1}$  & unreliability$^{2}$ & $z$ Cut & $x_{1}$ & Color & Outliers& Final$^{3}$\\
  \hline\noalign{\smallskip}
Cal\'an/Tololo& 28  &9      &0    & 0  &  0  & 0   & 19\\
CfA1&           21  &5      &6    & 0  &  1  & 0   & 9\\
CfA2&           24  &12     &2    & 0  &  0  & 0   & 10\\
CfA3&           87  &50     &0    & 1  &  2  & 0   & 34\\
CfA4&           74  &37     &2    & 0  &  0  & 1   & 34\\
CSP1&           21  &3      &3    & 2  &  1  & 0   & 12\\
CSP2&           40  &4      &4    & 0  &  3  & 1   & 28\\
LOSS&           150 &26     &21   & 0  &  8  & 0   & 95\\
all &           445 &146    &38   & 3  &  15 & 2   & 241\\
  \hline\noalign{\smallskip}
  \end{tabular}
\end{center}
  \begin{tablenotes}
  \item[*]{The number of SNe Ia removed by each
      selection criterion. Many SNe Ia fail multiple cuts.}
  \item[*]{$1$ The initial number of SNe Ia, after the removal of known peculiar
 SNe Ia, SNe Ia with clear photometric inconsistencies, and
 SNe Ia with better photometry from other samples.}
 \item[*]{$2$ SNe Ia with unreliable data due to an insufficient number
 of epochs or high Milky Way $E(B-V)$.}
 \item[*]{$3$ The number of SNe Ia satisfying all selection criterion.}
 \end{tablenotes}
 \label{sampletab}
\end{table*}

\subsection{Host galaxy properties}\label{sect:2.3}

\begin{figure}
   \centering
\includegraphics[width=90mm,angle=0]{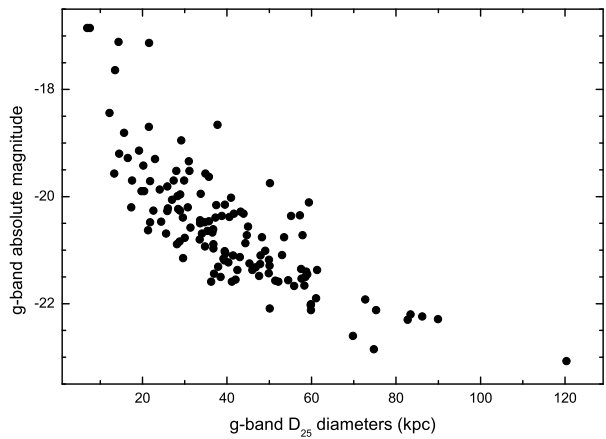}
\caption{%
The relation of galaxy size and absolute magnitude.
}
\label{dmagrel} %% no full stop at the end
\end{figure}

Our main host galaxy data are from \citep{2012A&A...544A..81H}.
We switch the g-band apparent magnitude of host galaxy into the absolute magnitude. K-correction is given by equation (\ref{Kcheck}), and the value of the second item is about 0.1 mag at z $\sim$ 0.1, about 0.05 mag at z $\sim$ 0.05. In our sample, only one galaxy has redshift above 0.1 and the majority (92.5\%) of these host galaxies have $\leq$ 0.05. Therefore we only consider the first K-correction item to compute absolute magnitude, e.g.,
\begin{equation}
\ K=2.5lg(1+z)+2.5lg\frac{\int^{\lambda_{2}}_{\lambda_{1}}I(\lambda)\phi_{\lambda}d\lambda}{\int^{\lambda_{2}}_{\lambda_{1}}
I(\frac{\lambda}{1+z})\phi_{\lambda}d\lambda}
\label{Kcheck}
\end{equation}
About galaxies sizes, we use the g-band diameters measured at $\mu_{g} = 25$ mag arcsec$^{-2}$ isophotal level.  In Fig.\ref{dmagrel}, we show the relation between the g-band size and absolute magnitude. Table \ref{galsampletab}
lists the properties of host galaxies in SN Ia samples.

\begin{table*}
\caption[]{The properties of host galaxies in SN Ia samples.}
  \label{Tab:poptical}
\begin{center}
\begin{tabular}{lcccccc}
  \hline\noalign{\smallskip}
Sample& SALT2 output$^{1}$   & star-formings & AGNs& passives & three types$^{2}$ & Diameter/magnitude$^{3}$\\
  \hline\noalign{\smallskip}
Cal\'an/Tololo& 19   & 0   &1    &0  & 1    &1    \\
CfA1&           9    & 0   &1    &2  & 3    &8    \\
CfA2&           10   & 0   &1    &1  & 2    &5    \\
CfA3&           34   & 5   &4    &6  & 15   &20   \\
CfA4&           34   & 2   &3    &5  & 10   &18   \\
CSP1&           12   & 3   &1    &2  & 6    &8    \\
CSP2&           28   & 4   &2    &4  & 10   &14   \\
LOSS&           95   & 7   &9    &11 & 27   &58   \\
all &           241  & 21  &22   &31 & 74   &132  \\
  \hline\noalign{\smallskip}
  \end{tabular}
\end{center}
  \begin{tablenotes}
  \item[*]{$1$ the SALT2 survival number of SNe Ia, after all kinds of
      selection criterion in the last section.}
 \item[*]{$2$ the sum of three types of galaxies: star-formings, AGNs and passives .}
 \item[*]{$3$ galaxies which have magnitude or size data.}
 \end{tablenotes}
 \label{galsampletab}
\end{table*}

To examine the dependence of SN Ia properties on environmental properties, we classify the SN Ia hosts into different groups. According to the active extent of host galaxies, we take the measures of
\citet{2012A&A...544A..81H} who used WHAN diagram to separate host galaxies into three subsamples: star-forming galaxies, Active Galactic Nucleis (AGNs) and passive galaxies.
The second split is performed according to their luminosity:
galaxies with g-band magnitude $\leq-20.75$ mag are classified as high luminosity, those with g-band magnitude $>-20.75$ mag as low luminosity. The third split is based on host size:
galaxies with $\mu_{g}=25$ mag arcsec$^{-2}$ diameters $\leq37.5$ kpc are small size, and others as large size.
The exact values chosen as the split points are a somewhat subjective choice.
The luminosity and size split points are both chosen to separate the hosts into bins of approximately equal
sizes. (We consider the effect of varying the last two split points
in later sections.)

\section{Correlations with SN fit parameters}\label{sect:3}

SALT2 reports an corrected B-band peak apparent magnitude ($m^{corr}_{B}$), a stretch
value ($x_{1}$) and a color (or $c$) term for each individual SN, they have a relation of
\begin{equation}
\ m^{corr}_{B} = m_{B} + \alpha x_{1} - \beta c,
\label{salt2}
\end{equation}
where $\alpha$ describes the overall stretch law for the sample, $\beta$ is the color law for the whole sample, $m_{B}$ and c are only corrected for Milky Way extinction without host galaxy extinction. $\alpha$ and $\beta$ are typically determined from simultaneous fits with the cosmological parameters.

\subsection{Host active extent}\label{sect:3.1}

\begin{figure}
   \centering
\includegraphics[width=\textwidth, angle=0]{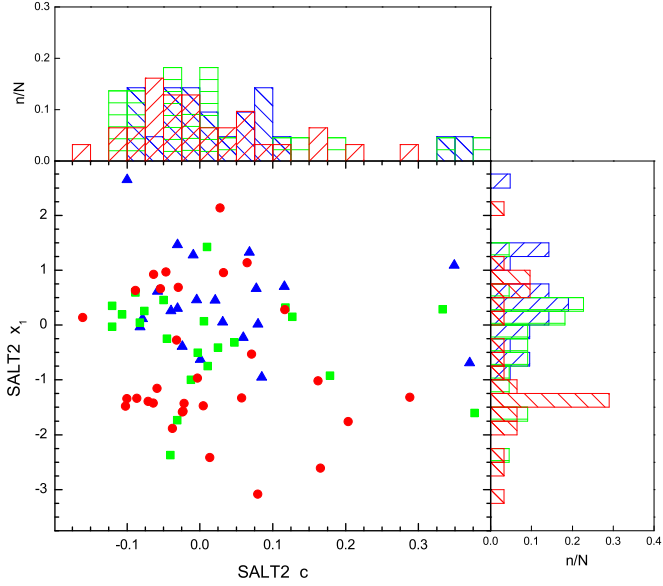}
\caption{%
The observed distribution of $x_1$ and $c$ (color) values for host active extent.
Blue solid triangular indicate SNe Ia in star-forming galaxies, green solid squares indicate SNe Ia in AGNs, and red solid circles denote SNe Ia in passive galaxies.
The histograms in the top panel of the figure show the normalized distribution of $c$ for star-forming galaxies (blue) , AGNs (green) and passive galaxies (red).
The right-hand panel shows the histograms of $x_1$ for different host galaxies.}
\label{sfrxc} %% no full stop at the end
\end{figure}

In Fig.\ref{sfrxc}, we show the SALT2 output ($x_1$ and $c$ values) according to host galaxy active extent. The solid (blue) triangulares denote SNe Ia in star-forming galaxies, the solid (green) squares indicate SNe Ia in AGNs, and the solid (red) circles show SNe Ia in passive galaxies. The distribution of $x_1$ is obviously different. In Table \ref{sfrxctab}, we show the mean values and
standard deviations about $x_1$ and $c$.
In agreement with the results given by \citet{2010ApJ...722..566L}, we confirm that SNe Ia present a clear difference in the $x_1$ distributions, and SNe Ia with small $x_1$ favor passive galaxies, while SNe Ia with large $x_1$ favor star-forming galaxies. Using t-test, we find that the mean value of $x_1$ in star-forming galaxies is significantly larger than which in other type galaxies,
with passive galaxies at a confidence of 4$\sigma$, and with AGNs at a confidence level of 2.7$\sigma$. However the mean value of $x_1$ is not obviously different in AGNs and passive galaxies.

From Fig.\ref{sfrxc}, we note that there is no relation between the color term ($c$) of SNe Ia and host galaxy types.  For t-test, at 2$\sigma$ confidence level, there is no significant difference for $c$ values in three type host galaxies, implying that the rest-frame colors of SNe Ia are dominated either by local, circum-stellar dust with the same color distributions, or by the same intrinsic color variations in all galaxy types.

\begin{table*}
\caption[]{Statistic values of SN fitting parameters for host galaxy types.}
  \label{sfrxctab}
\begin{center}
\begin{tabular}{lccccc}
  \hline\noalign{\smallskip}
galaxy type&     N& mean($x_{1}$)  &StdDev($x_{1}$)& mean($c$)& StdDev($c$) \\
  \hline\noalign{\smallskip}
star-formings&      21    & $0.404$    &0.842   &0.038  &0.123 \\
AGNs&               22    & $-0.261$   &0.865   &0.021  &0.134 \\
passives&           31    & $-0.737$   &1.265   &0.007  &0.010 \\
  \hline\noalign{\smallskip}
  \end{tabular}
\end{center}
\end{table*}

\subsection{Host luminosity}\label{sect:3.2}
\begin{figure}
   \centering
\includegraphics[width=\textwidth, angle=0]{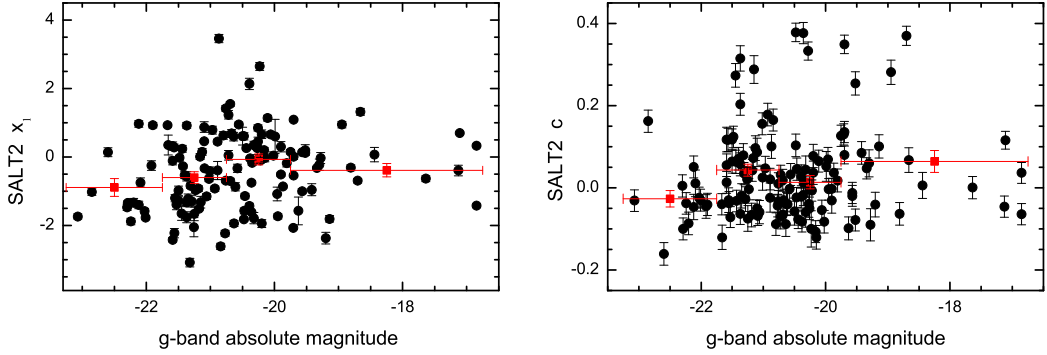}
\caption{%
The left-hand figure shows the dependence of observed $x_1$ on g-band absolute magnitudes.
The red squares bins stand for the mean value of $x_1$ in each bin.
The right-hand figure is similar with the left but now for the $c$.
}
\label{magxc} %% no full stop at the end
\end{figure}

In Fig.\ref{magxc}, we show the SALT2 output ($x_1$ and $c$) dependence on Host g-band absolute magnitudes.
At about 2$\sigma$ confidence level, we find
SNe Ia with smaller $x_1$ trend to host in high luminosity host galaxies.
However, at 2$\sigma$ confidence level, there is no correlation between host absolute magnitude and $c$.

\subsection{Host size}\label{sect:3.3}
\begin{figure}
\centering
\includegraphics[width=\textwidth, angle=0]{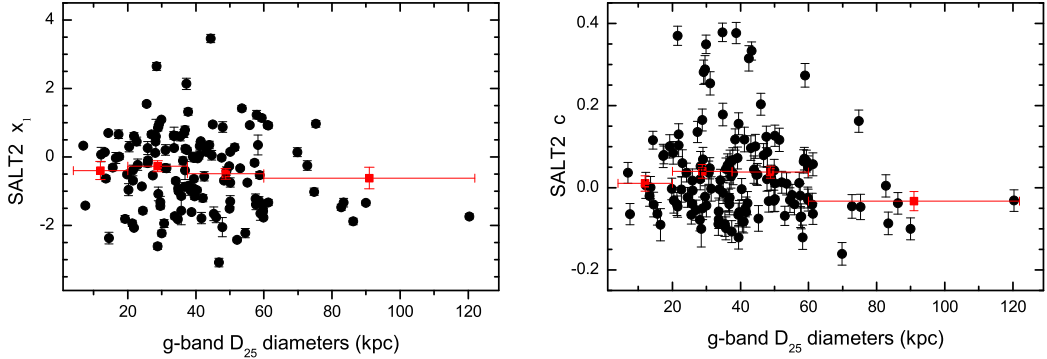}
\caption{%
As Fig.\ref{magxc}, but for  size instead of luminosity.}
\label{dxc} %% no full stop at the end
\end{figure}

In Fig.\ref{dxc}, we show the SALT2 output ($x_1$ and $c$) with different host size.
At 2$\sigma$ confidence level, there is no correlation between $x_1$ and host size,
and $c$ is also not correlated with host size.

\section{Residuals from global cosmological fits on host galaxies}\label{sect:4}

We now do $\chi^2$ fit by the equation:
\begin{equation}
\label{eq3}
\chi^2=\sum_{N}\frac{(m^{corr}_{B} - m^{mod}_{B}(z,\mathcal{M}_B
;\Omega_{M}))^{2}}{\sigma_{\mathrm{stat}}^2 + \sigma_{\mathrm{int}}^2},
\end{equation}
where $m^{corr}_{B}$ is given by equation (\ref{salt2}), $m^{mod}_{B}$ is the $B$-band
magnitude of cosmological model for each SN Ia given by
\begin{equation}
\label{eq4}
\ m^{mod}_{B}=5\log_{10}{\mathcal D_L}(z;\Omega_{M}) + \mathcal{M}_B,
\end{equation}
where ${\mathcal D_L}$ is reduced luminosity distance.
$\mathcal{M}_B=M_{B}+5\log_{10}(c/H_0)+25$, where $M_{B}$ is the absolute
magnitude of a SN Ia in the $B$-band. For SALT2 fitter, $M_{B}$ is given under
$x_{1}=0$ and $c=0$. For convenience, we present our results
as $M_{B}$ rather than $\mathcal{M}_B$, but it is noted that
H$_{0}$ is taken as 70 km s$^{-1}$ Mpc$^{-1}$.

$\sigma_{\mathrm{int}}$ parameterizes the intrinsic dispersion of each SN Ia,
and the sum is over $N$ SNe Ia entering the fit. $\sigma_{\mathrm{stat}}$ is the total identified statistical error and includes the uncertainties in $m_{B}$, $m^{mod}_{B}$ and peculiar velocity that is set as $400$ km/s in the paper.

One way is to examine the residuals of SNe Ia from cosmological fit using the entire
low redshift sample and $\Omega_{M}$ is set as 0.270. Throughout, we define
a Hubble residual $(HR)$ as $HR=m^{corr}_{B} - m^{mod}_{B}$, implying that after light-curve correction brighter SNe Ia have
negative Hubble residuals.

We use Alex Conley's $\tt minuit\textunderscore cosfitter$\footnote{http://casa.colorado.edu/$^{\sim}$aaconley/Software.html}
code to do $\chi^2$ cosmological fitting for the entire sample that has 241 SNe Ia.
We assume an intrinsic dispersion of $\sigma_{int}=0.12$ mag that adds the error on the distance modulus to make a reduced $\chi^2$ close to one (i.e., $\chi^2/{\rm ndf}\approx 1$). We get the nuisance
parameters of $(M_{B}, \alpha, \beta)$ = ($-19.080$, 0.148, 3.059), shown in Table \ref{cosmictab}.

\begin{table*}
 \caption{Best fitting values for $M_{B}$, $\alpha$, $\beta$ and $rms$.} %% no full stop at the end of caption
\begin{center}
\begin{tabular}{cccccc}
  \hline\noalign{\smallskip}
 $\alpha$ & $\beta$ & $M_{B}$ &$rms$ & N & $\chi^2$\\
  \hline\noalign{\smallskip}
  $0.148 \pm 0.012$ & $3.059 \pm 0.116$ & $-19.080 \pm 0.013$ &
0.196 & 241 & 240.2\\
  \hline\noalign{\smallskip}
  \end{tabular}
\end{center}
\label{cosmictab}
\end{table*}
We use the best-fit cosmological results of the entire sample to get new subsamples according to the host galaxy properties, such as host type submsaple (74 SNe Ia), host absolute magnitude submsaple (132 SNe Ia) and  host size subsample (132 SNe Ia).

\subsection{Host active extent}\label{sect:4.1}

\begin{table*}
\caption[]{Statistic values of Hubble Residuals for host-galaxy types.}
  \label{Tab:poptical}
\begin{center}
\begin{tabular}{lccc}
  \hline\noalign{\smallskip}
galaxy type&        N&        mean  &StdDev   \\
  \hline\noalign{\smallskip}
star-formings&      21    & $0.055$    &0.158   \\
AGNs&               22    & $0.024$    &0.172   \\
passives&           31    & $-$0.035   &0.186  \\
  \hline\noalign{\smallskip}
  \end{tabular}
\end{center}
\label{sfrrsetab}
\end{table*}

In Table \ref{sfrrsetab}, we show the residuals of different host galaxies. We make a t-test for the residuals of different host galaxies. At 2.1$\sigma$ confidence, the residuals in passive galaxies are more negative than those in star-forming galaxies.
However, at 2$\sigma$ confidence, there is no significant difference of the residuals between AGNs and any other galaxies.

\subsection{Host luminosity}\label{sect:4.2}

\begin{table*}
\caption[]{Statistic values of Hubble Residuals for host g-band absolute magnitude.}
  \label{Tab:poptical}
\begin{center}
\begin{tabular}{cccccccc}
  \hline\noalign{\smallskip}
luminosity split & \multicolumn{3}{c}{high luminosity hosts} & \multicolumn{3}{c}{low luminosity hosts}& significance$^{1}$\\
$M_{g}$& N & mean  &StdDev & N & mean  &StdDev &\\
\hline\noalign{\smallskip}
$-20.25$&   90 & $-0.026$ & 0.188 & 42 &0.095 &0.214 &3.2$\sigma$\\
$-20.5$&    73 & $-0.044$ & 0.187 & 59 &0.083 &0.205 &3.8$\sigma$\\
$-20.75$&   62 & $-0.055$ & 0.187 & 70 &0.072 &0.201 &3.9$\sigma$\\
$-21$&      52 & $-0.060$ & 0.184 & 80 &0.059 &0.204 &3.6$\sigma$\\
$-21.25$&   39 & $-0.077$ & 0.196 & 93 &0.050 &0.197 &3.5$\sigma$\\
  \hline\noalign{\smallskip}
  \end{tabular}
\end{center}
\label{magrestab}
  \begin{tablenotes}
  \item[*]{$1$  It shows the significance level where SNe Ia in high luminosity hosts are brighter than those in low luminosity hosts.}
 \end{tablenotes}
\end{table*}

In Table \ref{magrestab}, we show the residuals of two luminosity groups and the effect of different split points. At $>$3$\sigma$ significance level, the residuals in high luminosity hosts are more negative than those in low luminosity hosts, implying that SNe Ia in high luminosity galaxies are brighter than those in high luminosity hosts.

Here, we also fit for a linear dependence of residual on host g-band absolute magnitude or size using the package LINMIX \citep{2007ApJ...665.1489K}, which was used to determine the significance of trends with residuals by \citet{2010ApJ...715..743K}. LINMIX is a
Bayesian approach for linear regression using a Markov chain Monte Carlo (MCMC) analysis, assuming that the
measurement errors are Gaussian. We make the assumption that our errors on Hubble residuals are Gaussian
and input into LINMIX, and the average of the upper and lower 1$\sigma$ uncertainties is taken as the error in the dependent variable.

The residuals with host g-band absolute magnitudes are shown in the left-hand of Fig.\ref{dmagres}.
The overplotted lines are the best-fit model determined from LINMIX. In
all our LINMIX analysis, we use 10,000 MCMC realizations. For the residuals relating with absolute magnitudes, we obtain the best-fit relation

\begin{equation}
HR=(0.066 \pm 0.015) \times M_{g} - 1.413 \pm 0.290,
\label{eq5}
\end{equation}

The MCMC realizations in LINMIX are used to generate
a sampling of the posterior distribution on the slope. Of
the MCMC realizations, the slope is greater than
zero with nearly 100\%. Fitting a Gaussian to the posterior slope distribution, we yield a mean of 0.066 and a standard deviation
of 0.015. From the Gaussian fit, the mean slope does not equal zero at 4$\sigma$ confidence level, implying that there is a correlation between g-band absolute magnitude and the residual.

\subsection{Host size}\label{sect:4.3}

\begin{table*}
\caption[]{Statistic values of Hubble Residuals for host size.}
  \label{Tab:poptical}
\begin{center}
\begin{tabular}{cccccccc}
  \hline\noalign{\smallskip}
 size split & \multicolumn{3}{c}{large size hosts} & \multicolumn{3}{c}{small size hosts}& significance$^{1}$\\
$D_{25}$& N & mean  &StdDev & N & mean  &StdDev &\\
\hline\noalign{\smallskip}
30&     90 & $-0.010$ & 0.204 & 42 &0.058 &0.197 &2.1$\sigma$\\
35&     78 & $-0.024$ & 0.197 & 54 &0.065 &0.205 &2.7$\sigma$\\
37.5&   67 & $-0.015$ & 0.177 & 65 &0.040 &0.227 &2$\sigma$\\
40&     58 & $-0.025$ & 0.184 & 74 &0.042 &0.215 &2.2$\sigma$\\
45&     44 & $-0.031$ & 0.180 & 88 &0.034 &0.213 &2.1$\sigma$\\
  \hline\noalign{\smallskip}
  \end{tabular}
\end{center}
  \begin{tablenotes}
  \item[*]{$1$  It shows the significance level where SNe Ia in large size hosts are brighter than those in small size hosts.}
 \end{tablenotes}
\label{drestab}
\end{table*}

The statistic values of the residuals and the effect of the different split luminosity points are shown in Table \ref{drestab}.
At $\geq$2$\sigma$ confidence level, the residuals in large size hosts are more negative than those in small size hosts, implying that SNe Ia in large size galaxies are brighter.

\begin{figure}
\centering
\includegraphics[width=\textwidth, angle=0]{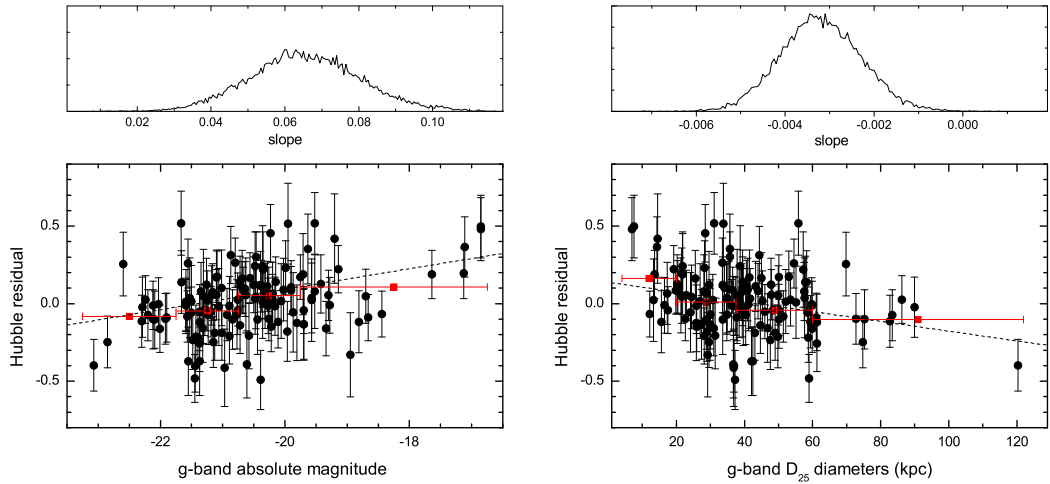}
\caption{%
The left-hand figure shows the residuals with g-band absolute magnitudes.
The right-hand panel shows the residuals with host diameters at $\mu_{g} = 25$ mag arcsec$^{-2}$.
The red square bins stand for the mean value of residuals in each bin. The overplotted line shows
the best fit to all data points as described in Section 4.2 and 4.3.}
\label{dmagres} %% no full stop at the end
\end{figure}

In the right-hand of Fig.\ref{dmagres}, we plot the residuals with host galaxy sizes and give a best-fit relation:
\begin{equation}
HR=(-0.0032 \pm 0.0009) \times D_{25} + 0.138 \pm 0.040,
\label{eq6}
\end{equation}
where $D_{25}$ stands for g-band size. Of the MCMC realizations, there is a negative
non-zero slope at 3.3$\sigma$ confidence level.

\section{Conclusions}\label{sect:5}

In this paper we have examined the photometric properties of SNe Ia in different host galaxies
using low redshift samples from literature.
We summaries the main conclusions of the paper as follows:

We confirm, to high significance, a strong correlation between host galaxy active extent type and the observed
width of the light curve, i.e., small $x_1$ favors passive host galaxies, while large $x_1$ favors star-forming galaxies. There is no significant different for $x_1$ in AGNs and other galaxies. No significant difference of $c$ appears in three type hosts. We find that, at about 2$\sigma$ confidence, smaller $x_1$ tends in
high luminosity hosts. However, at 2$\sigma$ confidence, host luminosity is not correlated with $c$,
neither is host size correlated with $x_1$ or $c$.

At 2.1$\sigma$ confidence, the residuals in passive galaxies are more negative than those in star-forming galaxy.
This result is consistent with that was found by \citet{2010ApJ...722..566L} at the median redshift. We infer that the correlation between SNe Ia luminosities and their host types could not vary with the redshift.
At $>$3$\sigma$ confidence level,
the residuals in high luminosity hosts are more negative.
We also find that, at $>$2$\sigma$ confidence level, SNe Ia have more negative residuals in large size hosts.

Using LINMIX, we find that there is a relation between the residual and host g-band absolute magnitude at 4$\sigma$ confidence level,
implying that over-luminous SNe Ia easily appear in high luminosity galaxies.
About the relation of the residual with host size,
the LINMIX fitting shows that there is a negative non-zero slope at 3.3$\sigma$ confidence level, which is higher than 2.6$\sigma$ given by  \citet{2010ApJ...715..743K}.

In a larger low redshift SN Ia sample, we find that SN Ia luminosity changes with host galaxies, involving
host type, luminosity and size.

 \normalem
\begin{acknowledgements}

We would like to thank Alex Conley and Mark Sullivan for
their meaningful helps during the course of this research.
We thank the CfA Supernova Group, the
Carnegie Supernova Project (CSP) and other authors to provide SN Ia data.
This research has also used the NASA/IPAC Extragalactic Database (NED)
and SIMBAD Astronomical Database. We also thank Hakobyan for providing host galaxies data.
We acknowledge the financial supports from the National Basic Research Program of China (973 Program 2009CB824800), the National Natural Science Foundation of China 11133006, 11163006, 11173054, and the Policy Research Program of Chinese Academy of Sciences (KJCX2-YW-T24).
\end{acknowledgements}

\appendix


\begin{thebibliography}{28}

\bibitem[Conley et al.(2011)]{2011ApJS..192....1C} Conley, A., Guy, J.,
Sullivan, M., et al.\ 2011, \apjs, 192, 1


\bibitem[Conley et al.(2008)]{2008ApJ...681..482C} Conley, A., Sullivan,
M., Hsiao, E.~Y., et al.\ 2008, \apj, 681, 482


\bibitem[Contreras et al.(2010)]{2010AJ....139..519C} Contreras, C., Hamuy,
M., Phillips, M.~M., et al.\ 2010, \aj, 139, 519


\bibitem[Ganeshalingam et al.(2010)]{2010ApJS..190..418G} Ganeshalingam,
M., Li, W., Filippenko, A.~V., et al.\ 2010, \apjs, 190, 418


\bibitem[Gupta et al.(2011)]{2011ApJ...740...92G} Gupta, R.~R., D'Andrea,
C.~B., Sako, M., et al.\ 2011, \apj, 740, 92


\bibitem[Guy et
al.(2007)]{2007A&A...466...11G} Guy, J., Astier, P., Baumont, S., et al.\ 2007, \aap, 466, 11


\bibitem[Guy et
al.(2010)]{2010A&A...523A...7G} Guy, J., Sullivan, M., Conley, A., et al.\ 2010, \aap, 523, A7


\bibitem[Hakobyan et
al.(2012)]{2012A&A...544A..81H} Hakobyan, A.~A., Adibekyan, V.~Z., Aramyan, L.~S., et al.\ 2012, \aap, 544, A81


\bibitem[Hamuy et al.(1996)]{1996AJ....112.2408H} Hamuy, M., Phillips,
M.~M., Suntzeff, N.~B., et al.\ 1996, \aj, 112, 2408


\bibitem[Hamuy et al.(2000)]{2000AJ....120.1479H} Hamuy, M., Trager, S.~C.,
Pinto, P.~A., et al.\ 2000, \aj, 120, 1479


\bibitem[Hicken et al.(2009)]{2009ApJ...700..331H} Hicken, M., Challis, P.,
Jha, S., et al.\ 2009, \apj, 700, 331


\bibitem[Hicken et al.(2012)]{2012ApJS..200...12H} Hicken, M., Challis, P.,
Kirshner, R.~P., et al.\ 2012, \apjs, 200, 12


\bibitem[Hicken et al.(2009)]{2009ApJ...700.1097H} Hicken, M., Wood-Vasey,
W.~M., Blondin, S., et al.\ 2009, \apj, 700, 1097


\bibitem[Hillebrandt
\& Niemeyer(2000)]{2000ARA&A..38..191H} Hillebrandt, W., \& Niemeyer, J.~C.\ 2000, \araa, 38, 191


\bibitem[Jha et al.(2006)]{2006AJ....131..527J} Jha, S., Kirshner, R.~P.,
Challis, P., et al.\ 2006, \aj, 131, 527


\bibitem[Jha et al.(2007)]{2007ApJ...659..122J} Jha, S., Riess, A.~G.,
\& Kirshner, R.~P.\ 2007, \apj, 659, 122


\bibitem[Kelly(2007)]{2007ApJ...665.1489K} Kelly, B.~C.\ 2007, \apj, 665,
1489


\bibitem[Kelly et al.(2010)]{2010ApJ...715..743K} Kelly, P.~L., Hicken, M.,
Burke, D.~L., Mandel, K.~S., \& Kirshner, R.~P.\ 2010, \apj, 715, 743


\bibitem[Kessler et al.(2009)]{2009ApJS..185...32K} Kessler, R., Becker,
A.~C., Cinabro, D., et al.\ 2009, \apjs, 185, 32


\bibitem[Lampeitl et al.(2010)]{2010ApJ...722..566L} Lampeitl, H., Smith,
M., Nichol, R.~C., et al.\ 2010, \apj, 722, 566


\bibitem[Landolt(1992)]{1992AJ....104..340L} Landolt, A.~U.\ 1992, \aj,
104, 340


\bibitem[Perlmutter et al.(1999)]{1999ApJ...517..565P} Perlmutter, S.,
Aldering, G., Goldhaber, G., et al.\ 1999, \apj, 517, 565


\bibitem[Riess et al.(1998)]{1998AJ....116.1009R} Riess, A.~G., Filippenko,
A.~V., Challis, P., et al.\ 1998, \aj, 116, 1009


\bibitem[Riess et al.(1999)]{1999AJ....117..707R} Riess, A.~G., Kirshner,
R.~P., Schmidt, B.~P., et al.\ 1999, \aj, 117, 707


\bibitem[Stritzinger et al.(2011)]{2011AJ....142..156S} Stritzinger, M.~D.,
Phillips, M.~M., Boldt, L.~N., et al.\ 2011, \aj, 142, 156


\bibitem[Sullivan et al.(2010)]{2010MNRAS.406..782S} Sullivan, M., Conley,
A., Howell, D.~A., et al.\ 2010, \mnras, 406, 782


\bibitem[Suzuki et al.(2012)]{2012ApJ...746...85S} Suzuki, N., Rubin, D.,
Lidman, C., et al.\ 2012, \apj, 746, 85


\bibitem[Wang
\& Han(2012)]{2012NewAR..56..122W} Wang, B., \& Han, Z.\ 2012, \nar, 56, 122


\end{thebibliography}
\end{document}